\documentstyle[prl,multicol,aps,epsf]{revtex}

\begin{document}

\draft


\title{Suppresion of Bloch oscillations by weak disorder in 
semiconductor superlattices}

\author{Enrique Diez,$^{1,3}$
Francisco\ Dom\'{\i}nguez-Adame,$^{2,3}$
and Angel S\'{a}nchez$^{1,3}$}

\address{$^1$Departamento de Matem\'aticas,
Escuela Polit\'ecnica Superior,\\
Universidad Carlos III, E-28911 Legan\'{e}s, Madrid, Spain\\
$^2$Departamento de F\'{\i}sica de Materiales,
Facultad de F\'{\i}sicas,\\
Universidad Complutense, E-28040 Madrid, Spain\\
$^3$Grupo Interdisciplinar de Sistemas
Complicados, Escuela Polit\'ecnica Superior,\\
Universidad Carlos III, E-28911 Legan\'{e}s, Madrid, Spain}

\date{\today}

\maketitle

\begin{abstract}

We investigate the dephasing dynamics of Bloch oscillations in
semiconductor superlattices by means of a very simple model including
disorder and applied electric fields. A thorough numerical study of our
model allows us to claim that small, unintentional well width fluctuations 
can be responsible for fast dephasing of Bloch oscillations 
at low temperatures. We show that the lifetime of Bloch oscillations 
is controlled by a characteristic time which depends
on the degree of disorder and is independent of the electric field. 
This result is further supported by the excellent agreement
between our model calculations and several recent experiments,
and leads to specific new predictions.

\end{abstract}


\begin{multicols}{2}

\narrowtext

Dynamical effects in quantum-well semiconductors superlattices (SL's) are
the basis for designing ultra-high speed electronic devices, as 
have been recently proposed \cite{Bouchard1}.  This idea of semiconductor SL's
operating at terahertz frequencies was already suggested a long time ago
by Esaki and Tsu \cite{Esaki}, who argued that electrons should undergo
periodic Bloch oscillations (BO's) \cite{Bloch}:  Under an applied electric
field $F$, provided that the interband coupling is negligible,
electrons oscillate in real space as well as in $k$ space with a
characteristic period given by $\tau_{\text B}=2\pi \hbar/eFd$, $d$
being the spatial period of the SL \cite{Dignam,Fernando}.  The
amplitude of BO's in real space is $A=\Delta/2eF$, where $\Delta$ is the
minibandwidth.  The coherent carrier motion is thus restricted to a region
of length $2\,A$. This periodic motion persists until the Bloch
electron loses energy gained from the field through scattering processes.
Reports of unambiguous experimental evidences for BO's in
GaAs-Ga$_{1-x}$Al$_{x}$As are presently available
\cite{Leo,Feldmann,Waschke,Plessen,Leisching}.

Inelastic
scattering by phonons, deviations from SL's perfect periodicity due to
unintentional imperfections, intraband scattering, interminiband transitions,
and scattering by impurities severely
reduce the quantum coherence required for the observation of BO's. 
Indeed, the scattering time $\tau$ must be 
larger than the Bloch period $\tau_{\text B}$ and therefore the electric
field must be larger than certain critical electric field $F_c$
\cite{Plessen}.  However, even in the most favorable experimental conditions
$\tau$ is not much larger than $\tau_{\text B}$ and thus only a few
BO's are usually observed. 
The origin of such loss of quantum
coherence in actual devices is far from understood
and, at present, there is much debate about the role played by different
scattering mechanisms in those processes. In this regard, Plessen {\em
et al.\/} \cite{Plessen} found that quantum coherence is lost after few
BO's in $30\,$\AA\ GaAs/$30\,$\AA\ Ga$_{0.7}$Al$_{0.3}$As SL's, which
was attributed to scattering by LO phonons. On the other hand, theoretical
studies point
out that under most experimental conditions interminiband transitions
are negligible and, consequently, cannot be responsible for the signal
decay \cite{Bouchard1}.  Furthermore, Plessen {\em et al.\/}
\cite{Plessen} conclude from their experimental results that the
critical electric field $F_c$ is higher for SL's with $\Delta$ larger
than the energy of LO phonons, $E_{\text{LO}}=36\,$meV.  They explain
this dependence by assuming that LO phonon emission is excluded when
$\Delta < E_{\text{LO}}$.  On the contrary, Leisching {\em et al.\/}
\cite{Leisching} detected up to six BO's but they did not observe any sign
of a phonon threshold in SL's with $\Delta$ ranging from $13$ up to
$46\,$meV.  These authors argued that the reduced sample quality of
Ref.~\onlinecite{Plessen} could be the responsible for the threshold.

From the above discussions, it becomes clear that understanding
the interplay between the electric field and
the imperfections of the SL's is crucial to elucidate the
discrepancies among different groups, either to pinpoint its relevance
or to exclude it. As far
as we know, however, a complete study of the effects of interface
roughness on the Bloch oscillations dynamics is currently lacking.
In this letter, we introduce a
theoretical model for imperfect GaAs-Ga$_{1-x}$Al$_{x}$As SL's that
successfully accounts for the experimental results.  We study the
dynamical behavior of these disordered SL's subject to a dc
electric field 
by measuring the position of the centroid of the wavepacket and by means of 
the  time dependent inverse participation ratio (IPR), to be
defined below.  These quantities will allow us 
to conclude that the assumption of weak disorder is enough to 
explain all the available
experimental data, thus firmly connecting the dephasing of BO's to the
quality of the sample. 

Interface roughness appearing during 
growth in {\em actual} SL's depend 
critically on the growth conditions\cite{Mader}. 
For instance, protrusions of one 
semiconductor into the other cause in-plane disorder and break translational 
invariance parallel to the layers. To  describe local excess or defect of 
monolayers, we allow the quantum well widths to fluctuate uniformly
around the nominal values; this can be seen as substituting
the nominal sharp width by an {\em average} along the parallel plane of
the interface imperfections.
Our approximation is valid whenever
the mean-free-path of electrons is much smaller than the in-plane
average size of protrusions as electrons only {\em see} {\em
micro\/}-quantum-wells with small area and uniform thickness \cite{Mader}.
In each {\em micro\/}-quantum-well presents a slightly different
value of its thickness and, as a consequence, resonant coupling between
electronic states of neighboring GaAs layers is decreased.  
Therefore, in the following we will take the width of the $n$th 
quantum well as $a(1+W\epsilon_n)$, where $W$ is a positive parameter
measuring the maximum fluctuation, $\epsilon_n$'s are distributed according 
to a uniform probability distribution, $P(\epsilon_n)=1$ if $|\epsilon_n|<1/2$ 
and zero otherwise,  $a$ is the nominal quantum well width.
For clarity we assume that the barrier width $b$ is the same in the whole SL,
although we have checked that this assumption can be dropped without
changing our conclusions.

For our present purposes, it is enough to focus on electron states
close to the conduction-band edge and use the effective-mass approximation.
The envelope-functions for the electron wavepacket satisfies the following
quantum evolution equation
\begin{equation}
i\hbar\,\frac{\partial\Psi(x,t)}{\partial t} = \left[
-\,{\hbar^2\over 2m^{*}}\,{d^2\phantom{x}\over dx^2} +
V(x) - eFx \right]\,\Psi(x,t),
\label{1}
\end{equation}
where $x$ is the coordinate in the growth direction and $V$ is
the SL potential at
flatband.  We have considered a constant effective-mass $m^{*}$ at the
$\Gamma$ valley for simplicity, but our numerical results should
qualitatively describe actual SL's with position-dependent
effective-mass.

We study the quantum dynamics of an initial Gaussian wavepacket
\begin{equation}
\Psi(x,0) = \left[2 \pi \sigma^2\right]^{-1/4}\,
\exp{\left[\frac{ik_0x-(x-x_0)^2}{4\sigma^2}\right]},
\label{3}
\end{equation}
where the mean kinetic energy is $\langle E\rangle=\hbar^2k^2_0/2m^{*}$
and $\sigma$ measures the width of the electron wavepacket. We stress
that, according to Bouchard and Luban \cite{Bouchard1} the dynamical
behavior of this initial state is similar to that of more realistic
functions.  The solution of Eq.~(\ref{1}) is accurately obtained using
the Cayley's form for the finite difference representation of its formal
solution \cite{Bouchard2,NuevoPRB}. Once the solution is obtained, 
we evaluate the position of the centroid of the wavepacket as
\begin{equation}
X(t) = \int_{-\infty}^{\infty}dx\,(x-x_0)|\Psi(x,t)|^2,
\label{3b}  
\end{equation}
which should display BO's. Moreover, we use the time-dependent inverse
participation ratio (IPR), defined as the second moment of the
probability density
\begin{equation}
\mbox{IPR}(t) = \int_{-\infty}^{\infty}dx\,|\Psi(x,t)|^4,
\label{4}
\end{equation}
to evaluate the spatial extent and the degree of localization of
electronic wavepackets.  We note that delocalized states 
present small IPR (in the ballistic limit, without applied field,
it vanishes with time  as $t^{-1}$), while localized states have larger IPR.

We have considered the same parameters as those of the SL's used in previous
experiments \cite{Plessen,Leisching}. In particular,
we present here results for the first one of these SL's, i.e., $100$ periods
of  $30\,$\AA\ GaAs and $30\,$\AA\ Ga$_{0.7}$Al$_{0.3}$As\cite{Plessen}. 
Samples are labeled according to their period length $d=a+b$,
namely $60$\AA\ SL.  Similar results are obtained
with the other SL's like the $84$, $97$, or $128\,$\AA\
($b=17\,$\AA , $a=67$, $80$ and $111\,$\AA, respectively),  i.e., the ones
reported by Leisching {\em et al.\/} \cite{Leisching},
although we do not present here these results for brevity.
We have straightforwardly calculated the miniband-width for the
 $60$\AA\ SL obtaining $\Delta = 90 \,$meV, being larger than $E_{\text{LO}}$.
We study applied  electric field in the 
ranges from $5$ up to $20\,$kV/cm.  The fluctuation parameter runs
from $W=0$ (perfect SL's) up to $W=0.20$ (strongly disordered SL's).
This maximum value considered here represents excess or
defect of a few monolayers. This value is above the degree of
perfection now achievable with MBE, so that realistic results are
comprised within this range and we do not need to analyze stronger disorder
values. 

\begin{figure}
\setlength{\epsfxsize}{6.0cm}
\centerline{\mbox{\epsffile{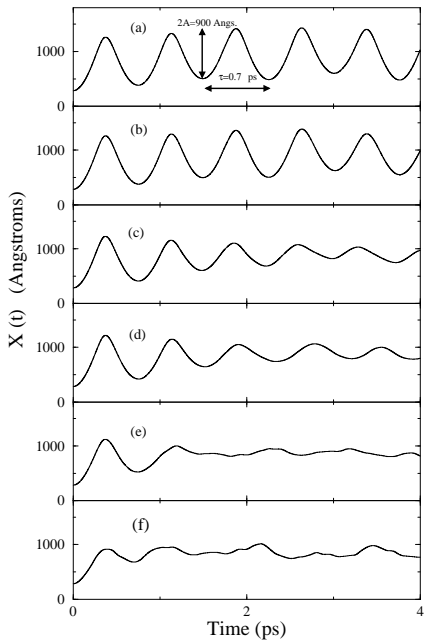}}}
\vspace*{1.0cm}
\caption{Centroid of an initial Gaussian wavepacket with
$k_0 = 0\,$ and $\sigma=300\,$\AA\
as a function of time in $60\,$\AA\ SL's. The applied electric
field is $F=10\,$kV/cm. From top to bottom (a) $W=0$, (b) $0.01$, (c) $0.03$,
(d) $0.05$, (e) $0.10$, and (f) $0.20$. The values of the amplitude
$2A=900\,$\AA\ and Bloch period $\tau=0.7\,$ps., for the perfect SL's (a),
are in excellent agreement with the theoretical predictions.}
\label{fig0}
\end{figure}

\begin{figure}
\setlength{\epsfxsize}{6.5cm}
\centerline{\mbox{\epsffile{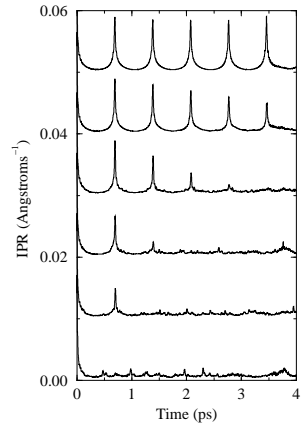}}}
\vspace*{-1.0cm}
\caption{IPR {\em vs} time for an initial Gaussian wavepacket with
$k_0 = 0\,$ and $\sigma=20\,$\AA, subject
to an electric field $F=10\,$kV/cm in $60\,$\AA\ SL's. From top to bottom
$W=0$, $0.01$, $0.03$, $0.05$, $0.10$, and $0.20$.} 
\label{fig1}
\end{figure}

\begin{figure}
\setlength{\epsfxsize}{5.0cm}
\centerline{\mbox{\epsffile{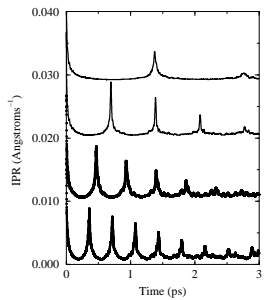}}}
\vspace*{-1.0cm}
\caption{IPR {\em vs} time for an initial Gaussian wavepacket with the
same parameters as in Fig.~2, placed in a $60\,$\AA\ SL's with $W=0.03$.
From top to bottom $F=5$, $10$, $15$, and $20\,$kV/cm.}
\label{fig2}
\end{figure}

Figure~\ref{fig0} displays the centroid position of the wavepacket
in the $60\,$\AA\ SL for $F=10\,$kV/cm and different values of the 
unintentional disorder. The initial Gaussian wavepacket is located 
in the centermost quantum well with $\sigma=300\,$\AA\ and $k_0 = 0\,$. 
In Fig.~\ref{fig0}(a), for an ordered SL, we observe  the occurrence
of very well defined BO's with amplitude $2A=900\,$\AA\ and 
period $\tau_{\text B} = 0.7 \,$ps, in excellent agreement with the 
theoretical predictions $2A=900\,$\AA\ and  $\tau_{\text B}=0.69\,$ps.
Notice, however, that the perfect oscillatory 
pattern detected in periodic SL's (upper panel) is progressively destroyed 
upon increasing the degree of disorder as seen in the rest of panels in
Fig.~\ref{fig0}, for $W=0.01$, $0.03$, $0.05$, $0.10$, and $0.20$.
It is most important to mention here that the results do not depend
on the particular realization of disorder.
We note that those values correspond, if we assume that a monolayer
width of this type of SL's is about $3$\AA , to a maximum 
excess of defect of less than one monolayer
($W=0.01$ and $0.03$), one monolayer ($W=0.05$), two monolayers ($W=0.10$) and
four monolayers ($W=0.20$).
The disorder induces a  decrease of the amplitude
of the oscillations and, besides, it  produces a progressive dephasing comparing
with the ideal perfect case [Fig~\ref{fig0}(a)].
In the strong disorder case no signs of BO's are found.  This
fact can be explained by the absence of translational invariance at
flatband and, consequently, by scattering of electrons with the random
potential.
Similar results are obtained with the SL's reported by Leisching {\em et al.\/}
\cite{Leisching}.
We have to note in this respect that for those samples 
we have observed
a slower decay of BO's, the slowest evolution corresponding to the SL with
the narrowest miniband.
This can be understood by noting that the amplitude of BO's is
proportional to the minibandwidth: 
 Thus, the coherent motion of electrons takes 
place in a smaller region when decreasing the mini-band-width and,
therefore, carriers are less influenced by disorder. This result is
in very good agreement with the experimental observations\cite{Leisching}.
\vspace*{1.5cm}
We can achieve a better resolution of the BO's period and
 the influence of the disorder by means of the IPR.
The upper panel of Fig.~\ref{fig1} presents the results for the IPR of
the $60\,$\AA\ SL when the initial Gaussian wavepacket is located in the
centermost quantum-well with $\sigma=20\,$\AA\ and $k_0$ = 0.
The electric field is $F=10\,$kV/cm.
In the absence of imperfections, the IPR displays a periodic
pattern with marked peaks at times $t_n=n\tau_{\text B}$, where $n$ is
any arbitrary, nonnegative integer and $\tau_{\text B} = 0.7 \,$ps.  This
means that the initial localized state is recovered after this time.
It is most important, to assess the
accuracy of our calculation,  to mention that the numerical
value of the IPR at maxima is exactly the same than that obtained from
(\ref{3}) and (\ref{4}), that is, $\mbox{IPR}(0) = 1/(2\sqrt{\pi}\sigma) =
0.01\,$\AA$^{-1}$.  Results corresponding to disordered SL's with the same
initial conditions as before are shown in the remaining panels of
Fig.~\ref{fig1},
confirming that BO's progressively disappear on increasing the degree of
disorder.  
\vspace*{-1.0cm}
{}From the above results we are led to the conclusion that there exists
a characteristic scattering time $\tau_{\text{dis}}$ after which BO's are
destroyed by disorder. Moreover, it is readily observed in
Fig.~\ref{fig1} that $\tau_{\text{dis}}$ decreases upon increasing the
degree of disorder.  However, the above results have been obtained for a
fixed value of the electric field, but clearly a meaningful definition
of the scattering time should be independent of the value of the
electric field.  To check the validity of the introduced
$\tau_{\text{dis}}$ we have studied the IPR for different values of the
applied electric field at a given degree of disorder.  Representative
results are presented in Fig.~\ref{fig2} for $W=0.03$ 
(on average less than one monolayer) and $F=5$, $10$, $15$ and
$20\,$kV/cm.  From this plot we can estimate that $\tau_{\text{dis}}
\simeq 2.5\,$ps for all values of the electric field.  Thus, this
scattering time plays the same role as the scattering time arising from
inelastic interactions, in the sense that $\tau_{\text{B}}$ must be kept
smaller than $\tau_{\text{dis}}$ to observe BO's. Interestingly, this value 
is the same as that obtained in the experiments of
Plessen {\em et al.\/} \cite{Plessen}
The scattering time increases when the minibandwidth decreases, for the
same amount of disorder, and values obtained with our model turn out to be
perfectly consistent with all the experimental 
values\cite{Feldmann,Leisching,Dekorsy}.
To conclude, we have been able to firmly connect BO's suppression and
dephasing in actual SL's to small deviations from exact flatness at
well-barrier interfaces. Specifically we have shown that an average
degree of imperfection of less than a monolayer suffices to explain
quantitatively the experimental results in \cite{Plessen,Leisching}.
Whereas the initially localized state is recovered 
after time $\tau_{\text{B}}$ in the case of perfect ($W=0$) SL's 
(regular behavior), 
any degree of disorder due to imperfections during growth
leads to the disappearance of BO's after a few oscillations: The higher the
degree of disorder the faster the vanishing of BO's. 
The very  good agreement with previous experiments points out the
crucial  role of imperfections in the dynamics of actual SLs
driven by electric fields.  Most importantly, we have been able to define a
characteristic scattering time $\tau_{\text{dis}}$, {\em independent} of
the electric field, after which BO's cannot be detected, this
being a specific prediction of our model that can be checked in
experiments. In other words,
for the BO's to be observed in actual SLs, the applied electric field 
must be larger than some critical electric field given by $eF_{\text{dis}}d =
2\pi\hbar/ \tau_{\text{dis}}$.  The existence of such a critical field is
evidently very important from the viewpoint of practical applications of our
results. $F_{\text{dis}}$ is directly related to the degree of
disorder present in the sample and
decreases upon increasing the quality of the sample,
i.e., it is an excellent parameter
to asses the performance of epitaxial growth techniques.

As we have seen, the main conclusion of this work, that
the importance of disorder in the transport properties of SL's has
to be underlined, in contrast with the general belief than the high
quality of actual SL's allows to neglect disorder as a second order effect.
It has to be kept in mind
that such an average of disorder of less than a monolayer  is currently 
unavoidable, more so
when preparing such long SL's ($100$ periods) as we have considered.
We note, however, that high-frequency
operating devices demand higher electric fields.  Therefore, for
sufficiently high fields, the region where coherent carrier motion takes
place, namely $2A=\Delta/eF$, is comparable to the SL's period $d$.  In
such a situation, the in-plane disorder is no longer well described by
an ensemble of different quantum wells as we have proposed 
because the wavepacket only
would {\em see} one quantum well in our model.  
Therefore more theoretical work is needed to investigate the role of
imperfections and other dephasing mechanisms like excitonic effects\cite{icps}
in the design of future, shorter period devices. 

The authors thank with great pleasure collaborative 
and illuminating conversations
with Fernando Agu\-ll\'o-Rueda, Karl Leo and Gintaras Valu\v{s}is.
This work has been supported by CICYT (Spain) under project MAT95-0325.

\end{multicols}

\end{document}